\documentclass[11pt]{article}
\usepackage[utf8]{inputenc}
\usepackage[a4paper]{geometry}
\usepackage{amsmath,amsbsy,amsgen,amscd,amsthm,amsfonts,amssymb,mathrsfs}

\usepackage{natbib}
\setcitestyle{authoryear,round,semicolon,aysep={,}}

\usepackage[dvipsnames]{xcolor}
\usepackage[colorlinks=true, linkcolor=blue, breaklinks=true]{hyperref}
\hypersetup{urlcolor=cyan}

\usepackage{algorithm}
\usepackage{algorithmic}
\usepackage{graphicx}  
\usepackage{caption}
\usepackage{subcaption}
\graphicspath{ {./images_paper/} }
\usepackage{multirow}
\usepackage{pgfplots}

\usepackage{titlesec}
\titleformat*{\section}{\large\bfseries}
\titleformat*{\subsection}{\normalsize\bfseries}
\titleformat*{\subsubsection}{\small\bfseries}

\theoremstyle{plain}
\newtheorem{theorem}{Theorem}[section]

\theoremstyle{definition}
\newtheorem{remark}[theorem]{Remark}

\usetikzlibrary{shapes.misc}
\tikzset{cross/.style={cross out, draw=black, minimum size=2*(#1-\pgflinewidth), inner sep=0pt, outer sep=0pt},
cross/.default={1pt}}

\newcommand{\SO}[1]{\operatorname{SO}(#1)}

\usepackage{booktabs}

\newcommand{\FW}{F_{W_1}}
\newcommand{\Fwemd}{F_{\text{\tiny WEMD}}}
\newcommand{\surr}{f_t}
\newcommand{\RBO}{R_{\text{est}}}
\newcommand{\RNM}{R_{\text{refine}}}
\newcommand{\Rtrue}{R_*}
\newcommand{\vtrue}{v_*}

\title{Alignment of Density Maps in Wasserstein Distance}
\author{Amit Singer and Ruiyi Yang} 
\date{Princeton University}

\begin{document}

\maketitle
{\let\thefootnote\relax\footnote{Contact:
\texttt{ry8311@princeton.edu}}}
\begin{abstract}
    In this paper we propose an algorithm for aligning three-dimensional objects when represented as density maps, motivated by applications in cryogenic electron microscopy. The algorithm is based on minimizing the 1-Wasserstein distance between the density maps after a rigid transformation. The induced loss function enjoys a more benign landscape than its Euclidean counterpart and Bayesian optimization is employed for computation. Numerical experiments show improved accuracy and efficiency over existing algorithms on the alignment of real protein molecules. In the context of aligning heterogeneous pairs, we illustrate a potential need for new distance functions. 
\end{abstract}

\section{Introduction}

Alignment of three-dimensional objects is an important task in applications ranging from computer vision and robotics such as shape registration and model retrieval \citep{saupe20013d,makadia2006fully,chen20063d}, to medical imaging and molecular biology where protein structures need to be aligned before further processing and conformational analysis \citep{kawabata2008multiple,joseph2020comparing,han2021vesper}. 
Given a pair of 3D objects which are rigid transformations of each other, the goal is to recover the relative translation and rotation that would match the two objects. As the alignment procedure often needs to be applied multiple times in the applications above, designing an accurate and efficient algorithm is of great significance.

In this paper, we shall be interested in the case where the 3D objects are represented as density maps, motivated by applications in cryogenic electron microscopy (cryo-EM) \citep{lawson2016emdatabank}.  
To formalize our setup, suppose $\phi_1, \phi_2:\mathbb{R}^3\rightarrow \mathbb{R}$ are two probability density functions representing the volumes, with $\phi_2$ being a transformed version of $\phi_1$, i.e., 
\begin{align}
    \phi_2(x)=\phi_1\big(\Rtrue(x+\vtrue)\big), \qquad \forall \,x\in\mathbb{R}^3, \label{eq:phi1 phi2}
\end{align}
for some $\vtrue\in\mathbb{R}^3$ and $\Rtrue\in \SO3$, the rotation matrix group. The goal of the alignment problem is to recover the rotation $\Rtrue$ and translation $\vtrue$ based on the density maps $\phi_1$ and $\phi_2$. Here we have assumed $\phi_1$ and $\phi_2$ to be probability densities only for framing our problem in Wasserstein distances below, while the proposed algorithm will work for density maps taking negative values or having non-unit masses. In practice, the volumes are given as three-dimensional arrays $V_i\in\mathbb{R}^{L\times L\times L}$ with $L$ an integer, which can be treated as discretizations of the $\phi_i$'s, with the voxel values encoding their configurations. 

A natural idea for solving the alignment problem is to search for the optimal translation and rotation through the following optimization task:
\begin{align}
    (\widehat{v},\widehat{R})
    \in \underset{(v,R)\in\mathbb{B}\times \SO3}{\operatorname{arg\,min}} \,\,d\big(\phi_1(R(\cdot+v)),\phi_2(\cdot)\big)=:\underset{(v,R)\in\mathbb{B}\times \SO3}{\operatorname{arg\,min}} \,\, F_d(v,R), \label{eq:alignment problem}
\end{align}
where $\mathbb{B}$ is a cube containing $\vtrue$ and $d$ is a suitable distance function on the space $\mathcal{P}(\mathbb{R}^3)$ of all probability measures over $\mathbb{R}^3$.
Most existing works with few exceptions (see Section \ref{sec:liter review}) set $d$ as the usual $L^2$ distance and \eqref{eq:alignment problem} is then solved by gradient-based methods or a type of exhaustive search over the space $\mathbb{B}\times \SO3$. 
However, due to the irregular shapes of the volumes, the landscape of $F_{L^2}$ could be highly nonconvex and gradient-based methods would fail with poor initialization. Exhaustive search-based methods, on the other hand, could return more accurate results but have formidable costs if implemented naively. Methods exploiting convolution structures of $F_{L^2}$ \citep{kyatkin2000algorithms} can lead to great computational speed up but are still considered expensive for large volumes.

Motivated by these issues, in this paper we shall propose an alignment algorithm based on solving \eqref{eq:alignment problem} in the $1$-Wasserstein distance, which is known to better reflect rigid transformations than Euclidean distances and hence creates a better loss landscape. 
Exploiting this fact, we employ tools from Bayesian optimization for numerical minimization of \eqref{eq:alignment problem}, which is able to return global optimizers yet with much fewer evaluations of the objective than an exhaustive search. 
The resulting algorithm achieves improved performance over existing ones as we will demonstrate on the alignment of real protein molecules.

\subsection{Wasserstein versus Euclidean Landscapes}\label{sec:landscape}
The main motivation for considering Wasserstein distances in \eqref{eq:alignment problem} comes from the better resulting loss landscape as we will discuss in this subsection. 
Recall that for two probability measures $\mu,\nu\in\mathcal{P}(\mathbb{R}^3)$, the $p$-Wasserstein distance for $p\in[1,\infty)$ is defined as 
\begin{align*}
    W_p(\mu,\nu) = \left(\underset{\gamma\in\Gamma(\mu,\nu)}{\operatorname{inf}}  \int_{\mathbb{R}^3\times \mathbb{R}^3} |x-y|^p d\gamma(x,y)\right)^{1/p},
\end{align*}
where $\Gamma(\mu,\nu)$ is the set of all couplings between $\mu$ and $\nu$, i.e., the set of all joint probability measures over $\mathbb{R}^3\times\mathbb{R}^3$ whose marginals are $\mu$ and $\nu$. Wasserstein distances have been widely studied and employed in for instance image retrieval \citep{rubner2000earth}, deep learning \citep{arjovsky2017wasserstein}, structural determination of molecular conformation \citep{zelesko2020earthmover} among many other areas of applied sciences.

For the alignment problem that we are interested in, the Wasserstein distances are better able to reflect the distances between a density map and its transformed version. For instance, it is shown by \cite[Lemma 3.5]{hamm2022wassmap} that for $p\in (1,\infty)$, 
\begin{align*}
    W_p\big(\phi(\cdot),\phi(\cdot+v)\big) = |v|, \qquad \forall \,v\in\mathbb{R}^3,
\end{align*}
where $\phi(\cdot+v)$ denotes the $v$-shifted density and $|\cdot|$ is the Euclidean norm. Therefore if the volumes are simply translations of each other (i.e. when $R$ equals identity $I_3$ in \eqref{eq:phi1 phi2}), then the associated loss in \eqref{eq:alignment problem} satisfies $F_{W_p}(v,I_3)=|v|$,  a convex function with a unique minimum at $v=0$. 
However, the same is far from being true for the $L^p$ loss $\|\phi(\cdot)-\phi(\cdot+v)\|_p$ if $\phi$ has an irregular shape.  

\begin{figure}[!htb]
    \centering  
    \minipage{0.2\linewidth}   
    \includegraphics[width=\linewidth]{25892.png}
    \vspace{-20pt}
    \subcaption{}\label{fig:landscape-vol}
    \endminipage
    \minipage{0.2\linewidth}  
    \includegraphics[width=\linewidth]{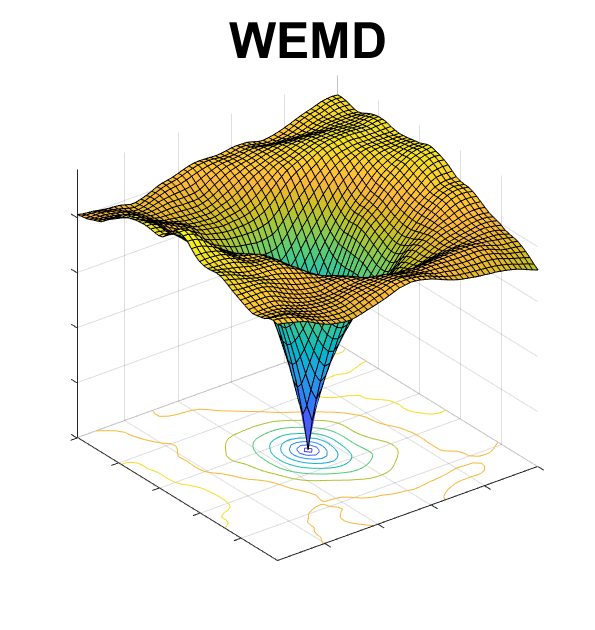}
    \vspace{-25pt}
    \subcaption{}\label{fig:landscape-wemd}
    \endminipage
    \minipage{0.2\linewidth}  
    \includegraphics[width=\linewidth]{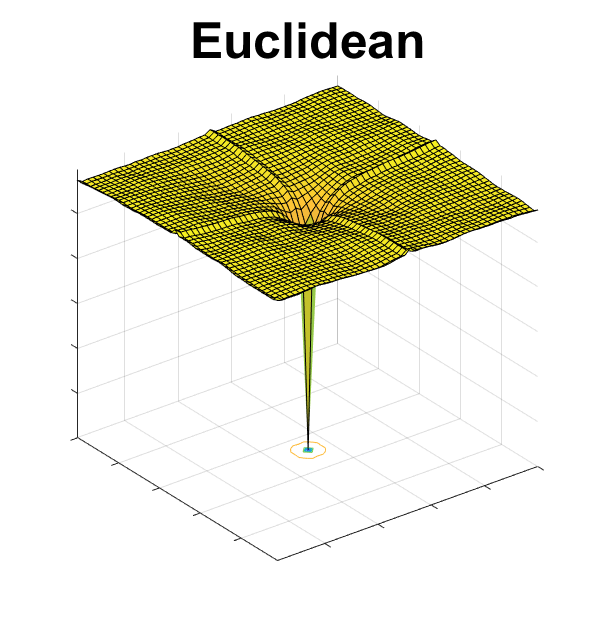}
    \vspace{-25pt}
    \subcaption{}\label{fig:landscape-eu}
    \endminipage
    \vspace{-0cm}
    \caption{  (a): visualization of the test volume.   (b)-(c): Comparison of local landscapes of $F_d(0,R)$ when $d$ is WEMD (cf. \eqref{eq:wemd}) and Euclidean ($L^2$)} \label{figure:landscape}
\end{figure}

Similar assertions can be made when the two volumes are pure rotations of each other (i.e. when $v=0$ in \eqref{eq:phi1 phi2}). Precisely, one can show that \cite[see e.g.][Propositions 1 and 2]{rao2020wasserstein}
\begin{align*}
    W_p\big(\phi(R_{\theta}\cdot),\phi(\cdot)\big)\leq 2\sin\left(\frac{\theta}{2}\right) M_p(\phi)^{1/p}
\end{align*}
for an in-plane rotation $R_{\theta}$ of angle $\theta$, where $M_p(\phi)$ is the $p$-th moment of $\phi$. The corresponding bound for the $L^p$ distance would have an additional factor of $\|\nabla \phi\|_{\infty}$, which could be large and gives a looser control on the change of $L^p$ distance with respect to the magnitude of $\theta$. 
Figures \ref{fig:landscape-wemd}-\ref{fig:landscape-eu} plot the loss $F_d(0,R)$ for two distance functions when $R=R_{\gamma}\cdot R_{\beta}$ represents a rotation around the $y$-axis by $\beta\in [-\pi/2,\pi/2]$ followed by a rotation around the $z-$axis by $\gamma\in [-\pi/2,\pi/2]$,   for the volume shown in Figure \ref{fig:landscape-vol}.   Here WEMD denotes the wavelet approximation of $W_1$ that we shall use for computation (see Section \ref{sec:wemd}) and Euclidean stands for the usual $L^2$ distance between vectors. 
We notice that the landscape of the loss associated with WEMD is flatter or has a larger basin of attraction compared with that for the Euclidean distance, which can facilitate the search of the minimizer. The narrow basin of attraction in the Euclidean case suggests the necessity of some type of exhaustive search unless the initial guess for gradient-based methods happens to fall in such region.

\begin{remark}
Despite the improved landscape offered by Wasserstein distances, there are certain issues that could arise when aligning a heterogeneous pairs of volumes that we shall discuss in Section \ref{sec:heter}. In particular, Wasserstein distances could be unrobust to perturbations of the volumes, which motivates the need for new distance functions.   \qed 
\end{remark}

\subsection{Computation via Bayesian Optimization}\label{sec:BO overview}
The landscape analysis above suggests potential benefits in using Wasserstein distances as the loss function in solving \eqref{eq:alignment problem}. However, a question remains for its numerical optimization as Wasserstein distances are less analytically tractable and more computationally costly. In particular, computing gradients of \eqref{eq:alignment problem} would be challenging both analytically and numerically, and an exhaustive search would require a huge computational budget due to the lack of a convolution structure. With these issues in mind, we shall instead adopt a Bayesian optimization approach, which does not require gradient information of the objective \eqref{eq:alignment problem} while being able to return accurate solutions with much fewer evaluations of  \eqref{eq:alignment problem} than an exhaustive search.

First of all, let's make the following simple observation that the relative translation can be recovered by centering the two density maps, so that the problem reduces to estimating the rotation $\Rtrue$ alone. Indeed, suppose without loss of generality that $\phi_1$ is already centered, i.e., $\int_{\mathbb{R}^3}x\phi_1(x)dx=0$. Then
\eqref{eq:phi1 phi2} implies after a change of variable that 
\begin{equation*}
    \int_{\mathbb{R}^3} x\phi_2(x)dx = \int_{\mathbb{R}^3} x\phi_1\big(\Rtrue(x+\vtrue)\big) dx = \Rtrue^{-1}\int_{\mathbb{R}^3}x\phi_1(x)dx -\vtrue=-\vtrue. 
\end{equation*}
Therefore one can recover $\vtrue$ by computing the center of mass of $\phi_2$ and the shifted volume $\widetilde{\phi}_2(x):=\phi_2(x-\vtrue)=\phi_1(\Rtrue x)$ is then a purely rotated version of $\phi_1$. 
This leads to a viable approach for estimating the shift vector between $\phi_1,\phi_2$ when they are noise free, as we shall demonstrate in Section \ref{sec:numerics-compare}. For the rest of this paper, we shall mainly focus on the rotational recovery, i.e., we assume 
\begin{align*}
    \phi_2(x) = \phi_1(\Rtrue x),\qquad \forall \,x\in\mathbb{R}^3
\end{align*}
and find the best rotation that minimizes 
\begin{align}\label{eq:alignment problem rot}
    \widehat{R}\in\underset{R\in \SO3}{\operatorname{arg\,min}} \,\,d\big(\phi_1(R(\cdot)),\phi_2(\cdot)\big)=:\underset{R\in \SO3}{\operatorname{arg\,min}} \,\, F_d(R),
\end{align}
This will be achieved with Bayesian optimization as we overview next.

On a high level, Bayesian optimization is an iterative procedure that searches for optimizer candidates by solving a sequence of surrogate problems instead of the original one \eqref{eq:alignment problem rot}.  At the $t$-th iteration, one collects the candidates $\{R_i\}_{i=1}^t$  picked so far 
together with the associated function values $\{F_d(R_i)\}_{i=1}^t$ to form using Bayesian techniques a surrogate function $f_t$ whose landscape resembles $F_d$ while being much cheaper to optimize. The $(t+1)$-th candidate is then chosen by solving the surrogate problem
\begin{align}
    R_{t+1}\in\underset{R\in \SO3}{\operatorname{arg\,min}} \,\,\surr(R), \label{eq:subproblem}
\end{align}
and is incorporated to the history $\{R_i,F_d(R_i)\}_{i=1}^t$ for updating $f_t$. 
After a total number of $T$ iterations the approximate solution to \eqref{eq:alignment problem rot} is returned as 
\begin{align*}
    \widehat{R}\in\underset{t=1,\ldots,T}{\operatorname{arg\,min}} \,\,F_d(R_t).
\end{align*}
Note that the whole procedure only requires access to function evaluations of $F_d$ but not its gradient. 

Intuitively speaking, the algorithm explores the search space $\SO3$ based on the landscapes of the surrogate functions $\surr$'s, which will approximate that of $F_d$ as $t$ increases but are much simpler to decode. 
As can be expected, the construction of $\surr$ is crucial for the algorithm to perform well. In this paper, we shall take $\surr$ as a Gaussian process interpolant (see Section \ref{sec:AF} for more details), which admits a simple analytic formula whose gradient is also available in closed form so that the surrogate problems \eqref{eq:subproblem} can be solved cheaply.  
Our numerical experiments in Section \ref{sec:numerics-BO} show that a total of $T=200$ iterations would suffice for accurate rotational alignment for real protein molecules, suggesting it as a practical algorithm.  Furthermore, the fact that only evaluations of $F_d$ are required implies that the proposed framework can be applied to arbitrary loss functions in \eqref{eq:alignment problem rot}, beyond vanilla Wasserstein or Euclidean distances. This could be a useful property when aligning a pair of heterogeneous volumes as we discuss in Section \ref{sec:heter} where more sophisticated loss functions may be needed. The algorithm can also be extended beyond the density map assumption to volumes represented for instance by point clouds as long as one can define a suitable loss function as in \eqref{eq:alignment problem rot}.

\subsection{Our Contributions}
The contributions of this paper are summarized as following:
\begin{itemize}
    \item We propose a novel algorithm for aligning three-dimensional objects (Section \ref{sec:method}), which achieves both improved accuracy and efficiency over existing methods on real datasets of protein molecules from cryo-EM (Section \ref{sec:simulation-volume}).  
    \item Our algorithmic framework can be extended beyond the 1-Wasserstein distance to arbitrary distance functions. 
    In the context of aligning a heterogeneous (similar but non-identical) pair of volumes, we show a potential need for new loss functions in which case our algorithm can be seamlessly incorporated (Section \ref{sec:heter}). 
\end{itemize}

\section{Related Work}\label{sec:liter review}
A classical approach for rotational alignment when the volumes are represented as point clouds is to perform principal component analysis and  align the resulting eigenvectors. This method could give exact recovery in theory (up to reflections) but is not robust to perturbations of the given volumes, which is usually the case in practice, and breaks when the volumes admit certain symmetry. More recent works in the computer vision literature include \cite{makadia2006fully,kazhdan2007approximate,althloothi2013robust} (see also the references therein), which again assume representations of the volumes as point clouds or other shape descriptors and parametrizations. Our problem setting is slightly different in that the point clouds representing our volumes are always fixed as the same Cartesian grid, where it is the voxel values that characterize their configurations. Nevertheless, we remark that the proposed algorithm in this paper can be extended easily to the point cloud setting. 

In the context of density map alignment that we consider, the optimization approach \eqref{eq:alignment problem} is most often adopted, which is then solved by gradient-based or exhaustive search-based methods. Setting $d$ as the $L^2$ distance, \eqref{eq:alignment problem} is equivalent to correlation maximization
\begin{align}
    (\widehat{v},\widehat{R})=\underset{(v,R)\in\mathbb{B}\times \SO3}{\operatorname{arg\,max}} \langle \phi_1(R(\cdot+v)), \phi_2(\cdot) \rangle_{L^2}, \label{eq:cross-corr}
\end{align}
where $\langle \cdot,\cdot\rangle_{L^2}$ denotes the $L^2$ inner product. 
The Chimera package \citep{pettersen2004ucsf} implements a steepest ascent algorithm for solving \eqref{eq:cross-corr} by relying on an initial alignment that is close to the true one, which is usually done manually by the user. 
\cite{chirikjian2004rotational} proposes using Kullback–Leibler divergence as the loss function which is later optimized with gradient descent, but no simulations are presented.  
  Setting $d$ as an entropic regularization of the 2-Wasserstein distance, the recent work \cite{riahi2023alignot} solves \eqref{eq:alignment problem} with stochastic gradient descent by iteratively computing the optimal transport plans. Improved results are obtained over Chimera but the algorithm still requires the initial alignment to be within certain range of the true one. Extending such ideas, \cite{riahi2023empot} uses Gromov-Wasserstein distance for partial alignment of density maps. These are the only works we are aware of that employ Wasserstein-based distances for the alignment problem. Our proposed method differs from \cite{riahi2023alignot,riahi2023empot} by employing Bayesian optimization and not involving explicit calculation of transport plans.

For this reason, exhaustive search-based methods are also popular and appear necessary. One subcategory of works in this direction attempts to search for the best alignment over a dense grid of translations and rotations that maximizes \eqref{eq:cross-corr}. 
Since the number of grid points is typically large, the key to these methods is a fast algorithm for computing \eqref{eq:cross-corr} given a pair of $v$ and $R$ \citep{kyatkin2000algorithms,chen2013fast,rangan2022radial}. 
This can be achieved by expanding the correlation in spherical harmonic bases (with the efficient spherical Fourier transform \citep{healy2002ffts}) and using the fact that rotation corresponds to application of Wigner-D matrices. 
Translations can be treated similarly with fast Fourier transform techniques. Existing packages include Xmipp \citep{de2013xmipp}, and EMAN2 \citep{tang2007eman2} which employs a hierarchical tree-based algorithm. 

Another subcategory of works approaches this problem by considering the projections of the volumes. The main idea is to not maximize the correlation between the volumes as in \eqref{eq:cross-corr} but instead their projections, where only inner products between images are computed and the search space can be reduced to five-dimensional \citep{yu2013projection}. 
The recent work \cite{harpaz2023three} improves this idea by employing common lines based techniques to further accelerate the search of matching projections.

\section{The Proposed Method}\label{sec:method}

In this section we present in detail our proposed method to solve \eqref{eq:alignment problem rot} with $d$ as the 1-Wasserstein distance:
\begin{align}\label{eq:alignment problem rot W}
    \widehat{R}\in\underset{R\in \SO3}{\operatorname{arg\,min}} \,\,W_1\big(\phi_1(R(\cdot)),\phi_2(\cdot)\big)=:\underset{R\in \SO3}{\operatorname{arg\,min}} \,\, \FW (R),
\end{align}    
Our algorithm is summarized in Algorithm \ref{algo:bo align}, which exploits the general framework of Bayesian optimization. To make our presentation self-contained, we shall introduce some necessary background before explicating the algorithmic details. We refer to \cite{frazier2018tutorial} for a more thorough introduction to Bayesian optimization. The interested reader can skip to Section \ref{sec:full algo} for a description of our full algorithm.  

As overviewed in Section \ref{sec:BO overview}, Bayesian optimization is an iterative procedure that searches for an optimizer of \eqref{eq:alignment problem rot W} by solving instead a sequence of surrogate problems. 
At each iteration, the surrogate problem is to optimize an \emph{acquisition function} constructed based on a \emph{probabilistic model} and all past queries of the objective function $\FW$. 
The acquisition function should be cheap to optimize while at the same time encodes enough information on the landscape of $\FW$. 
For our purpose in this paper, we shall adopt a \emph{Gaussian process} for probabilistic modeling of $\FW$ (Section \ref{sec:GP}) and a \emph{Gaussian process interpolant} for the acquisition function (Section \ref{sec:AF}).

\subsection{Gaussian Processes}\label{sec:GP}
The starting point of Bayesian optimization is to build a probabilistic model for the function $\FW$ in \eqref{eq:alignment problem rot W} that we wish to optimize, which is used to construct an acquisition function as in Section \ref{sec:AF}. Since $\FW$ is a nonparametric function, a common choice is to model it as a sample path from a Gaussian process \citep{rasmussen2006gaussian}. 
Recall that a Gaussian process (GP) over a space $\mathcal{M}$ is a collection of random variables $\{u(x),x\in\mathcal{M}\}$ where any finite subcollection is jointly Gaussian. For any finite set $\{x_i\}_{i=1}^N\subset \mathcal{M}$, the mean vector $m\in\mathbb{R}^n$ and the covariance matrix $\Sigma\in\mathbb{R}^{n\times n}$ of the finite-dimensional Gaussian $[u(x_1),\ldots,u(x_n)]^T$ can be specified in a consistent way through a mean function $\mu(\cdot)$ and a covariance function $c(\cdot,\cdot)$ so that  $m_i=\mu(x_i)$ and $\Sigma_{ij}=c(x_i,x_j)$.  Intuitively speaking, a GP is a random process whose realizations fluctuate around $\mu(\cdot)$ according to restrictions imposed by $c(\cdot,\cdot)$. In particular, $\mu(\cdot)$ and $c(\cdot,\cdot)$ completely determine the distribution of the GP and from the modeling perspective it suffices to make the appropriate choices for them.

Typically the mean is set to be zero and the covariance function encodes one's prior belief on sample path properties such as smoothness. When the space $\mathcal{M}$ is a subset of the Euclidean space, one of the most commonly used covariance functions is the squared exponential 
\begin{align}
    c(x,y)=\sigma^2 \exp\left(-\frac{|x-y|^2}{2\ell^2}\right),\qquad x,y\in\mathbb{R}^n, \label{eq:se cf}
\end{align}
where $\sigma,\ell>0$ are respectively the marginal variance and correlation lengthscale parameters. Roughly speaking, $\sigma$ determines the overall magnitude of the sample paths, and function values at two points with distance on the order of $\ell$ are nearly uncorrelated. An important feature of the squared exponential covariance function is that it leads to infinitely differentiable sample paths \cite[see e.g.][Section 4.2]{rasmussen2006gaussian}, which are suitable choices for modeling smooth functions.

Based on our discussion in Section \ref{sec:landscape} that the Wasserstein distances vary relatively smoothly with respect to rigid transformations, the squared exponential serves as a natural choice for our problem. Recall that in our setting the space $\mathcal{M}=\SO3$ is a subset of $\mathbb{R}^{3\times 3}$. 
Therefore we shall define our covariance function as 
\begin{align}
    c(R,S)=\sigma^2 \exp\left(-\frac{\|R-S\|_{F}^2}{2\ell^2}\right),\qquad R,S\in \SO3, \label{eq:cf eu}
\end{align}
where $\|\cdot\|_F$ denotes the Frobenius norm. Notice that this is equivalent to viewing each matrix as a vector in $\mathbb{R}^9$ and applying \eqref{eq:se cf}, as a result of which the covariance function \eqref{eq:cf eu} retains positive definiteness. Therefore the probabilistic model for $\FW$ in \eqref{eq:alignment problem rot W} that we shall adopt is a GP over $\SO3$ with mean zero and covariance \eqref{eq:cf eu}. 

\begin{remark}
Here we briefly discuss other possible choices of covariance functions over $\SO3$. On the one hand, any other Euclidean covariance functions such as the Mat\'ern family \cite[see e.g.][Section 4.2]{rasmussen2006gaussian} can be employed by viewing $\SO3$ as a subset of $\mathbb{R}^{3\times 3}$. Our choice of \eqref{eq:cf eu} is motivated by its superior empirical performance in our alignment problem and its simple form that facilitates numerical optimization of the acquisition function as will be discussed in Section \ref{sec:AF}.

On the other hand, the space $\SO3$ admits a manifold structure, which suggests a covariance function over the manifold $\SO3$ that takes geometry into account. 
This for instance can be achieved by 
\begin{align}
    c(R,S)=\sigma^2 \exp\left(-\frac{\Theta(R,S)^2}{2\ell^2}\right),\,\, \Theta(R,S)=\arccos\left(\frac{\operatorname{Trace}(RS^T)-1}{2}\right),\,\, R,S\in \SO3, \label{eq:cf geo}
\end{align}
where $\Theta(R,S)$ is the relative angle between the two rotations, whose absolute value is also the geodesic distance on $\SO3$. However, \emph{geodesic exponential kernels} such as \eqref{eq:cf geo} are not positive definite in general for all $\ell>0$, although empirical evidence suggests positive definiteness for a range of $\ell$'s in certain cases \citep{feragen2016open}.   Moreover, our simulation experience suggests slightly better accuracy when using \eqref{eq:cf eu} over \eqref{eq:cf geo} so we shall stick to the covariance function \eqref{eq:cf eu}.   Finally, we point to some more sophisticated covariance functions over $\SO3$ proposed by \cite{jaquier2022geometry}.  \qed 
\end{remark}

\subsection{Gaussian Process Interpolant as Surrogate}\label{sec:AF}
With a probabilistic model for $\FW$, we shall replace the original optimization problem \eqref{eq:alignment problem rot W} by a sequence of simpler surrogate problems. This will be achieved by constructing simple-to-optimize approximations $\surr$ to the objective function $\FW$ in \eqref{eq:alignment problem rot W}. In this paper, we shall take $\surr$ as the conditional expectation of the GP proposed in Section \ref{sec:GP} after ``observing'' the data $\{(R_i,\FW(R_i))\}_{i=1}^t$, which we explain now.

Suppose we have picked the first $t$ candidates $\{R_i\}_{i=1}^t$ (for the initial candidates, we can for instance generate $t_0$ random rotation matrices). Together with the associated objective function values $Y_i=\FW(R_i)$, we shall interpret the pairs $\{(R_i,Y_i)\}_{i=1}^t$ as observations we have obtained for the unknown function $\FW$. Now in the Bayesian regression framework, we have 
\begin{align*}
    \FW \sim \Pi, \qquad \FW (R_i)=Y_i,
\end{align*}
where $\Pi$ is the GP model as in Section \ref{sec:GP}. Therefore a natural estimator for $\FW$ is the conditional expectation
\begin{align*}
    \surr(x)=\mathbb{E}_{G\sim \Pi} \big[G(x)\,\big|\,G(R_i)=Y_i,1\leq i\leq t\big],
\end{align*}
which is known \citep{stein1999interpolation} to minimize the squared error loss $\mathbb{E}_{\FW\sim\Pi}|\FW(x)-\widehat{F}(x)|^2$ over all $\widehat{F}(x)$ that is measurable with respect to $\{\FW(R_i)\}_{i=1}^t$ when $\FW$ is indeed a sample path from $\Pi$. In particular, $\surr$ interpolates $\FW$, i.e., $\surr(R_i)=\FW(R_i)$ for $1\leq i\leq t$, 
and approximates $\FW$ increasingly well as more observations of $\FW$ are obtained. Furthermore, it can be shown that \cite[e.g.][Theorem 3.3] {kanagawa2018gaussian} $\surr$ admits a simple analytic formula 
\begin{align}
    \surr(x)=k(x)^T K^{-1}Y,\qquad x\in \SO3, \label{eq:AF}
\end{align}
where $k(x)\in \mathbb{R}^t$ is a vector with entries $[k(x)]_i=c(x,R_i)$, $K\in\mathbb{R}^{t\times t}$ is a matrix with entries $K_{ij}=c(R_i,R_j)$, and $Y\in\mathbb{R}^t$ is a vector with entries $Y_i=\FW(R_i)$. Notice that $\surr$ is simply a linear combination of the covariance functions 
\begin{align*}
    \surr(x)=\sum_{i=1}^t \big[K^{-1}Y\big]_i c(x,R_i),\qquad x\in \SO3,
\end{align*}
and admits an analytic formula for its Euclidean gradient under the covariance function choice \eqref{eq:cf eu} as
\begin{align}\label{eq:AF grad}
    \nabla^{\text{Eu}}\surr(x)=\sum_{i=1}^t \big[K^{-1}Y\big]_i c(x,R_i)\left(\frac{R_i-x}{\ell^2}\right), \qquad x\in \SO3.
\end{align}
Therefore optimization of $\surr$ can be carried out much more cheaply compared with the original problem \eqref{eq:alignment problem rot W} by supplying the gradient to standard optimization packages.

\subsection{The Full Algorithm}\label{sec:full algo}

Now we are ready to present the full algorithm, as summarized in Algorithm \ref{algo:bo align}. Starting with a GP model for $\FW$ and initial candidates $\{R_i\}_{i=1}^{t_0}$ (which can be randomly generated), we form the surrogate function \eqref{eq:AF} based on all observations $\{(R_i,\FW(R_i))\}_{i=1}^t$ obtained so far and search for the next candidate by solving
\begin{align}
    R_{t+1}\in\underset{R\in \SO3}{\operatorname{arg\,min}}\,\, \surr (R).  \label{eq:pick candiate}
\end{align}
Now with the new observation $(R_{t+1},\FW(R_{t+1}))$ included, we update the surrogate to $f_{t+1}$ and repeat the process \eqref{eq:pick candiate}.  
After a total of $T$ iterations, we shall return the candidate with smallest objective function value, i.e.,   
\begin{align}\label{eq:BO sol}
    \RBO\in \underset{t=1,\ldots,T}{\operatorname{arg\,min}} \,\,\FW(R_t)   
\end{align}
as the solution to the original problem \eqref{eq:alignment problem}. Such a procedure would serve as a prototype for our proposed method. To further improve its practicality, below we shall introduce two modifications: (i) approximation of the $W_1$ distance and (ii) an optional local refinement step.

\subsubsection{Wavelet Approximation of $W_1$ Distance}\label{sec:wemd}
Despite the favorable induced landscape as shown in Section \ref{sec:landscape}, a notable issue for Wasserstein distances is their high computational cost. For practical applications in cryo-EM that we shall consider, the density maps are represented as 
three-dimensional arrays of size $L\times L\times L$ with $L$ on the order of hundreds. Therefore the cost of exact computation of $W_1$ distances is prohibitive and scales in the worst case as $O(N^3\log N)$ with $N=L^3$. For this reason, we seek an approximation of $W_1$ through the wavelet approximation proposed by \cite{shirdhonkar2008approximate} which reduces the above cost to $O(N)$. 

Precisely, the wavelet earth mover's distance (WEMD) \citep{shirdhonkar2008approximate} is defined as 
\begin{align}
    \|\phi_1-\phi_2\|_{\text{\tiny WEMD}}=\sum_{\lambda} 2^{-j(1+n/2)} |\mathcal{W}\phi_1(\lambda)-\mathcal{W}\phi_2(\lambda)|, \label{eq:wemd}
\end{align}
where $n=3$ denotes the dimension of the density maps and $\mathcal{W}\phi_i$ denotes a 3D wavelet transform. The index $\lambda$ consists of the triplet $(\epsilon,j,k)$, where $\epsilon$ takes values in a finite set of size $2^n-1$ that for instance represents tensor products of 1D wavelets, and $j$ is the scale parameter that ranges over $\in\mathbb{Z}_{\geq 0}$ and $k$ ranges over $\in\mathbb{Z}^n$.  It is proved in \cite[Theorem 2]{shirdhonkar2008approximate} that the metric defined above is equivalent to the $W_1$ distance. Such an approximation has the additional advantage that the distance \eqref{eq:wemd} can be defined for density maps that take negative values, which is usually the case in practice, whereas the original $W_1$ distance is restricted to probability densities. 

Now we shall replace all occurrences of the $W_1$ distance in our previous algorithmic procedure by the WEMD distance \eqref{eq:wemd}. Precisely, the alignment problem we shall be solving becomes 
\begin{align}\label{eq:alignment problem rot wemd}
    \widehat{R}\in\underset{R\in \SO3}{\operatorname{arg\,min}} \,\,\|\phi_1(R(\cdot))-\phi_2(\cdot)\|_{\text{\tiny WEMD}}=:\underset{R\in \SO3}{\operatorname{arg\,min}} \,\, \Fwemd (R),
\end{align}
and the surrogate problem \eqref{eq:pick candiate} together with the returned solution \eqref{eq:BO sol} will be defined in terms of $\Fwemd$ instead. Note that the Bayesian optimization framework only requires access to function evaluations, so that replacing $\FW$ by $\Fwemd$ does not require extra modifications of the algorithm. In principle, $\Fwemd$ can be further replaced by any other distance function, which is an important feature of the adopted framework that we shall elaborate more in Remark \ref{remark:diff loss}.

\subsubsection{Local Refinement}
In Section \ref{sec:numerics-BO}, we will show numerically that after $T=200$ iterations, the algorithm described so far returns reasonably accurate recovery of the relative rotation $\Rtrue$ for real protein molecules. However, for a finite $t$, the surrogate $\surr$ only approximates $\FW$ and so does its minimizer. In order to obtain close-to-exact recovery, the candidates should form a dense enough cover of $\SO3$, which would require many more samples than $T=200$ and is computationally infeasible. For this reason, we introduce an optional local refinement step by employing the Nelder-Mead algorithm. Precisely, we shall return 
\begin{align}
    \RNM \in \underset{R\in \SO3}{\operatorname{arg\,min}}\,\, \|\phi_1(R(\cdot))-\phi_2(\cdot))\|_2,  \label{eq:refine sol}
\end{align}
where \eqref{eq:refine sol} is optimized with the Nelder-Mead algorithm initialized at $\RBO$ given by \eqref{eq:BO sol}. This concludes the description of our algorithm, presented in Algorithm \ref{algo:bo align}.

Note that we switched to the $L^2$ loss in \eqref{eq:refine sol} for reasons to be explained in Remark \ref{remark:refine L2}. We further remark that \eqref{eq:refine sol} can be solved in principle by other standard optimization algorithms such as BFGS. We have deliberately chosen Nelder-Mead because it only requires loss function evaluations in a similar spirit as Bayesian optimization. Such a property can be useful and crucial in the context of aligning heterogeneous pairs of volumes, where we show in Section \ref{sec:heter} a potential need for new and more sophisticated distance functions which one may only know how to evaluate. In this case our proposed framework would still be applicable.

\begin{algorithm}[!htb]
\caption{Volume Alignment in WEMD via Bayesian Optimization} \label{algo:bo align}
\begin{algorithmic}
\REQUIRE Volumes $\phi_1,\phi_2$; loss $\Fwemd$ \eqref{eq:alignment problem rot wemd};  GP covariance \eqref{eq:cf eu};  initialization $\{(R_i, \Fwemd(R_i))\}_{i=1}^{t_0}$.
\FOR{$t=t_0,\ldots,T$}
\STATE  Compute $\surr$ as in \eqref{eq:AF} and find $$R_{t+1}\in\underset{R\in \SO3}{\operatorname{arg\,min}}\,\, \surr(R)$$ 
\STATE  Add $(R_{t+1},\Fwemd(R_{t+1}))$ to $\{(R_i,\Fwemd(R_i))\}_{i=1}^t$. 
\ENDFOR
\STATE Set the estimated rotation as $$\RBO\in\underset{t=1,\ldots,T}{\operatorname{arg\,min}}\,\, \Fwemd (R_t).$$
\STATE (Optional) Solve the following with Nelder-Mead algorithm initialized at $\RBO$ 
$$\RNM \in \underset{R\in \SO3}{\operatorname{arg\,min}}\,\, \|\phi_1(R(\cdot))-\phi_2(\cdot))\|_2.$$
\ENSURE $\RBO$ and (optional) $\RNM$. 
\end{algorithmic}
\end{algorithm}

We end this section with further remarks on the algorithmic details.

\begin{remark}[Choice of $\surr$]\label{remark:gp-ucb}
In Bayesian optimization, $\surr$ is called the acquisition function and there have been extensive research on its choice \cite[see e.g.][]{frazier2018tutorial}, all of which could have been employed in our problem setting. Our choice of \eqref{eq:AF} is motivated by its simple form that admits an analytic formula for its gradient, which facilitates solving \eqref{eq:pick candiate}. For readers familiar with Bayesian optimization, \eqref{eq:AF} corresponds to GP-UCB \citep{srinivas2009gaussian} with no exploration, which appears to be a suboptimal choice. However for practical implementation, one may only want to solve \eqref{eq:pick candiate} approximately with early stopping to speed up the search, which can be treated as another form of exploration. Our experience suggests better performance for using \eqref{eq:AF} than adding a term proportional to the conditional variance as in the standard GP-UCB. Furthermore, \eqref{eq:AF} is independent of $\sigma$, the marginal variance parameter in \eqref{eq:cf eu}, which would otherwise be present in the standard form and requires tuning. \qed 
\end{remark}

\begin{remark}[Numerical optimization of $\surr$ and addressing handedness]\label{remark:manopt}
The majority of the efforts in Algorithm \ref{algo:bo align} are devoted to solving the surrogate problems \eqref{eq:pick candiate}, whose accuracy and efficiency are the key to the algorithm. We note that the kernel matrix $K$ in \eqref{eq:AF} could be ill-conditioned if two candidates $R_t$ and $R_{t'}$ are very close to each other. For numerical stability, a nugget term is usually incorporated to \eqref{eq:AF} so that we consider instead 
\begin{align}
    \surr=k(x)^T (K+\tau I_t)^{-1}Y,  \label{eq:AF tau}
\end{align}
with a small $\tau$ for practical implementation.  

Taking advantage of the manifold structure of $\SO3$, we shall optimize $\surr$ with the Riemannian optimization package \citep{manopt,townsend2016pymanopt}. 
The Euclidean gradient \eqref{eq:AF grad} can still be supplied for speed up, which is automatically transformed into Riemannian gradients by the package. In cryo-EM applications, one often needs to address also the handedness of the molecules, i.e., when $\phi_1$ and $\phi_2$ differ additionally by a reflection. 
This can be achieved in our framework by optimizing $\surr$ instead over $\operatorname{O}(3)$, the space of orthogonal matrices, which corresponds to the Stiefel manifold $\operatorname{St}(3,3)$ in the package \citep{manopt,townsend2016pymanopt}. However, we remark that this is not much different from reflecting one of $\phi_1,\phi_2$ first and aligning the resulting pair since the new search space $\operatorname{O}(3)$ is twice as large as $\SO3$ and the number of iterations $T$ for accurate alignment is also expected to double. 
\qed  
\end{remark}

\begin{remark}[Choice of loss function]\label{remark:diff loss}
Algorithm \ref{algo:bo align} is presented in terms of the loss function $\Fwemd$ \eqref{eq:alignment problem rot wemd} and only requires access to its evaluations but nothing else. An immediate observation is that Algorithm \ref{algo:bo align} can be applied to solve the alignment problem with any distance function $d$ in \eqref{eq:alignment problem rot}, as a consequence of the Bayesian optimization framework. Our choice of $W_1$ distance is motivated by the fact that it can be efficiently approximated with WEMD \eqref{eq:wemd}. Based on the discussion in Section \ref{sec:landscape}, we believe a general $W_p$ distance could also be used, with for instance an entropic regularization \citep{cuturi2013sinkhorn} for practical implementation. 

To illustrate the advantage of employing Wasserstein-based loss functions, we will show in Section \ref{sec:numerics-BO} the improved performance of Algorithm \ref{algo:bo align} over its $L^2$ loss counterpart.
Finally, in the context of aligning  a pair of heterogeneous volumes in Section \ref{sec:heter}, we shall discuss a potential need for new loss functions other than the vanilla Wasserstein or Euclidean distances. 
In such a setting, the new loss function could be analytically intractable and expensive to evaluate, which renders unclear the applicability of gradient-based or exhaustive search-based algorithms. Nevertheless, Algorithm \ref{algo:bo align} could be seamlessly incorporated as long as we can afford a small number of evaluations of the loss function and potentially give a feasible solution. \qed
\end{remark}

\begin{remark}[Refinement in $L^2$ distance] \label{remark:refine L2}
We have switched to the $L^2$ distance in the refinement step \eqref{eq:refine sol} for the following two reasons. First, the recovery $\RBO$ is already very close to the global minimizer of \eqref{eq:alignment problem rot wemd}, which is likely to lie also in the basin of attraction in the $L^2$ loss, although the latter is generally much narrower as shown in Figure \ref{figure:landscape}. 
Therefore a local search in $L^2$ loss would be sufficient and more efficient than in the WEMD.

Second, the WEMD or $W_1$ distance appear to be more sensitive than the $L^2$ distance to perturbations of the density maps in terms of the optimal alignment. 
In practice, the density maps may correspond to two different reconstructions of the same object and are only provided as three-dimensional arrays $V_1,V_2\in\mathbb{R}^{L\times L \times L}$ for an integer $L$. 
Therefore, the minimizers of the empirical versions of \eqref{eq:alignment problem rot wemd} and \eqref{eq:refine sol} are not necessarily equal to the true relative rotation $\Rtrue$ but only approximately. 
However, we found that the minimizer of the empirical version of \eqref{eq:alignment problem rot wemd} could be non-negligibly different from $\Rtrue$ in many cases where there is no issue with the $L^2$ distance, suggesting the $L^2$ loss as a more robust local search metric. 
Such observation is also related to the alignment of heterogeneous pairs that we shall discuss in Section \ref{sec:heter}. \qed
\end{remark}

\section{Numerical Experiments}\label{sec:simulation-volume}
In this section we apply our proposed method in Section \ref{sec:method} to align real protein molecules from a cryo-EM database \citep{lawson2016emdatabank}.
Section \ref{sec:implementation} contains the implementation details, in particular the choices of hyperparameters in Algorithm \ref{algo:bo align}. 
In Section \ref{sec:numerics-BO}, we investigate the performance of Algorithm \ref{algo:bo align} under different downsampling levels, total number of iterations, and noise corruption, while comparing with the $L^2$ loss version of Algorithm \ref{algo:bo align}. 
In Section \ref{sec:numerics-compare}, we compare the performance of Algorithm \ref{algo:bo align} with the two recent works \cite{harpaz2023three,riahi2023alignot}. The algorithmic complexity is discussed in Section \ref{sec:complexity}. Our code is available on \url{https://github.com/RuiyiYang/BOTalign}.

\subsection{Implementation Details}\label{sec:implementation}
For GP modeling, we shall use the Gaussian kernel defined in \eqref{eq:cf eu} with $\sigma=1$. Notice that the surrogate $\surr$ defined in \eqref{eq:AF} is independent of $\sigma$ so its choice is indeed arbitrary. The choice of $\ell$, on the other hand, would have an effect and is empirically tuned for optimal algorithmic performance. As mentioned, we shall investigate the performance of Algorithm \ref{algo:bo align} under both the WEMD loss and the $L^2$ loss. The values of $\ell$ are fixed as 0.75 and 1 respectively throughout the experiments. The WEMD distance is computed using PyWavelet \citep{lee2019pywavelets} with the \texttt{sym3} wavelet and maximum scale level $s=6$ following \cite{kileel2021manifold}.

We shall initialize Algorithm \ref{algo:bo align} with a single candidate $I_3$, the identity matrix. Our experience suggests that the initialization does not affect much the performance. 
For optimization of the surrogate problems, we shall follow the discussion in Remark \ref{remark:gp-ucb} and consider $\surr$ defined in \eqref{eq:AF tau} with $\tau=10^{-3}$ for numerical stability. 
The surrogate $\surr$ is then optimized with Riemannian steepest descent with random initialization using Pymanopt \citep{townsend2016pymanopt},
with an early stopping if both the gradient norm and the step size are less than 0.1. We empirically found that such an early stopping greatly improves the efficiency of Algorithm \ref{algo:bo align} while not losing much accuracy (see Remark \ref{remark:gp-ucb}). 

In practice, the density maps of the volumes are given as three-dimensional arrays $V\in\mathbb{R}^{L\times L\times L}$ for some integer $L$. In other words, $V$ is supported on the Cartesian grid and computing its rotated versions is a nontrivial and essential procedure. 
In our algorithm, this is done with the ASPIRE package \citep{garrett_wright_2023_7871783}, which first computes the nonuniform Fourier transform of $V$ over the rotated grid and then applies an inverse Fourier transform. Note that this step is needed when computing the loss function values in Algorithm \ref{algo:bo align}.

Finally, to further speed up the computation, a common practice in cryo-EM is to downsample the given volumes $V\in\mathbb{R}^{L\times L\times L}$ to be of size $\mathbb{R}^{L_0\times L_0\times L_0}$ for some integer $L_0<L$. This leads to faster computation of WEMD distances and potentially a better loss landscape by removing the fine-scale structures of the protein molecules. For the rest of this section, we shall treat $L_0$ and the total number of iterations $T$ in Algorithm \ref{algo:bo align} as user-chosen parameters and in Section \ref{sec:numerics-BO} demonstrate its performance when $L_0\in\{32,64\}$ and $T\in\{150,200\}$. 

\subsection{Alignment of Real Protein Molecules}\label{sec:numerics-BO}
\begin{figure}[!htb]
\centering
\minipage{0.125\textwidth}
\subcaption*{\scriptsize EMD-3683}
\includegraphics[width=\textwidth]{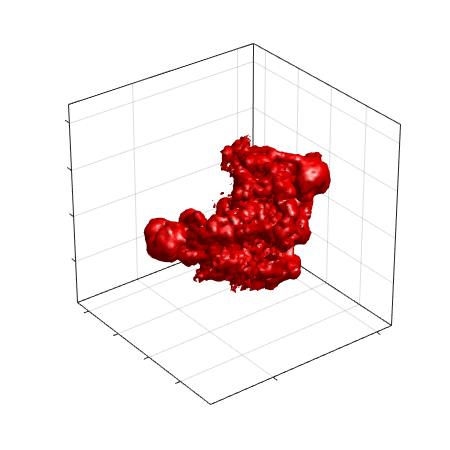}
\vspace{-0.65cm}
\subcaption*{\scriptsize $L=128$}
\endminipage
\minipage{0.125\textwidth}
\subcaption*{\scriptsize EMD-1717}
\includegraphics[width=\textwidth]{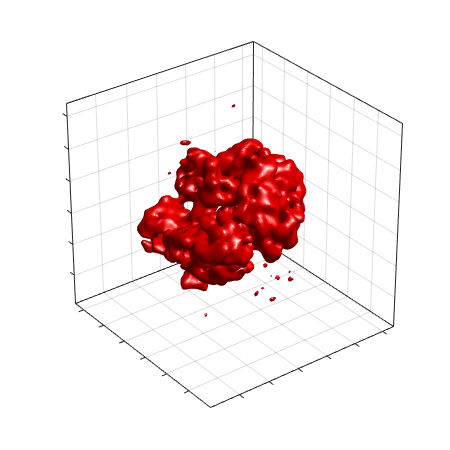}
\vspace{-0.65cm}
\subcaption*{\scriptsize $L=128$}
\endminipage
\minipage{0.125\textwidth}
\subcaption*{\scriptsize EMD-3342}
\includegraphics[width=\textwidth]{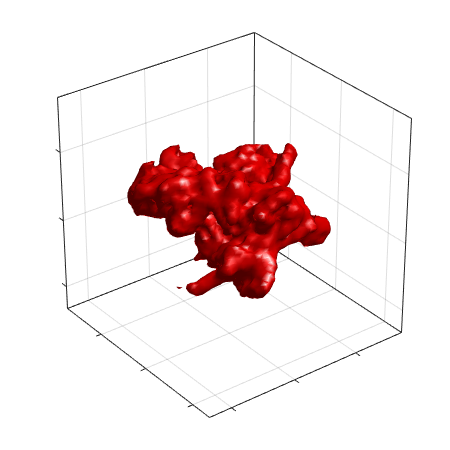}
\vspace{-0.65cm}
\subcaption*{\scriptsize $L=256$}
\endminipage
\minipage{0.125\textwidth}
\subcaption*{\scriptsize EMD-9515}
\includegraphics[width=\textwidth]{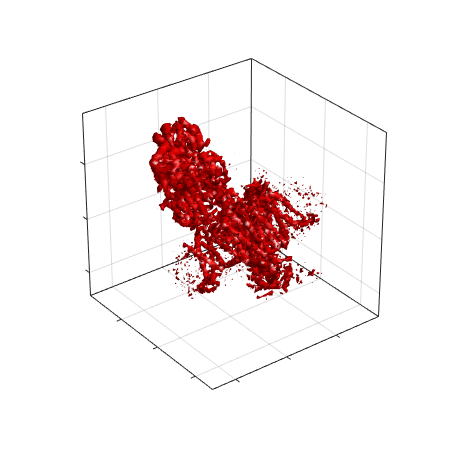}
\vspace{-0.65cm}
\subcaption*{\scriptsize $L=256$}
\endminipage
\minipage{0.125\textwidth}
\subcaption*{\scriptsize EMD-4547}
\includegraphics[width=\textwidth]{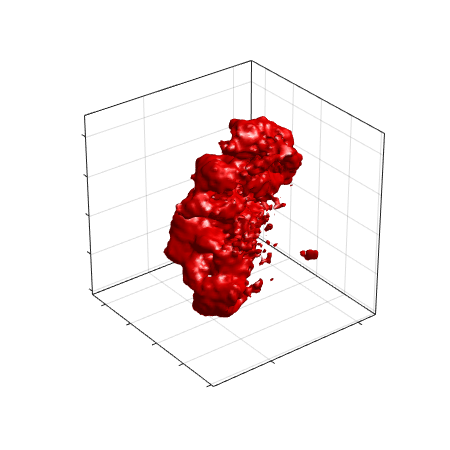}
\vspace{-0.65cm}
\subcaption*{\scriptsize $L=280$}
\endminipage
\minipage{0.125\textwidth}
\subcaption*{\scriptsize EMD-10180}
\includegraphics[width=\textwidth]{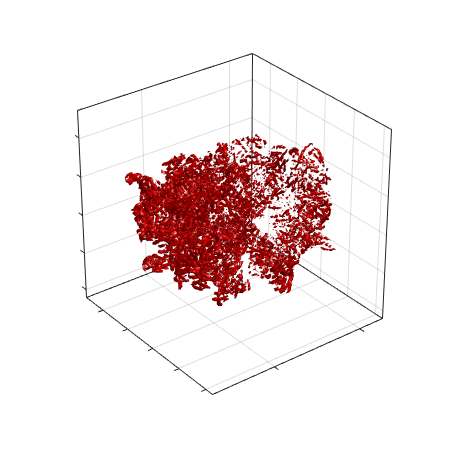}
\vspace{-0.65cm}
\subcaption*{\scriptsize $L=320$}
\endminipage
\minipage{0.125\textwidth}
\subcaption*{\scriptsize EMD-25892}
\includegraphics[width=\textwidth]{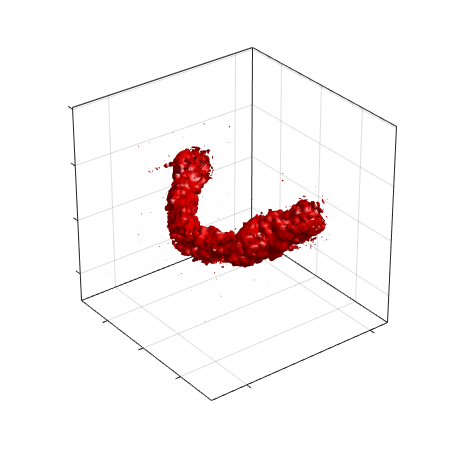}
\vspace{-0.65cm}
\subcaption*{\scriptsize $L=320$}
\endminipage
\minipage{0.125\textwidth}
\subcaption*{\scriptsize EMD-2660}
\includegraphics[width=\textwidth]{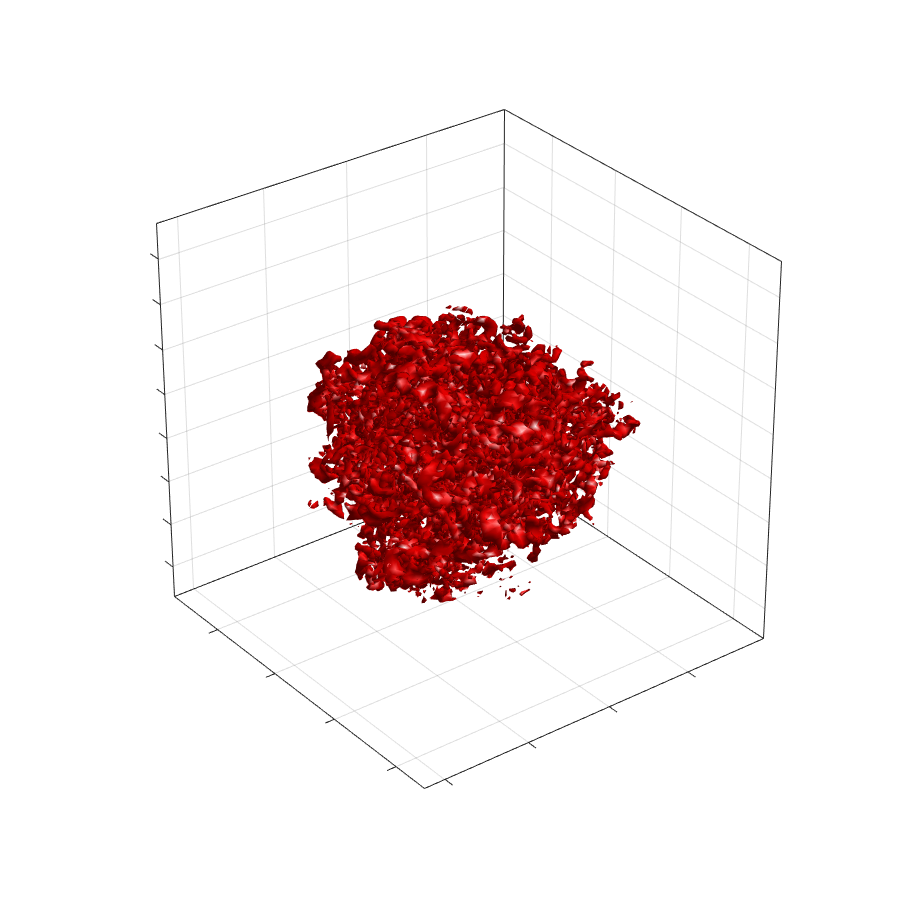}
\vspace{-0.65cm}
\subcaption*{\scriptsize $L=360$}
\endminipage
\caption{Visualization of the test volumes}\label{figure:volumes}
\end{figure}

In this section we shall illustrate the performance of Algorithm \ref{algo:bo align} on real protein molecules from the publicly available Electron Microscopy Data Bank \citep{lawson2016emdatabank}. The experimental setup is as follows. For a given volume $V_1\in\mathbb{R}^{L\times L\times L}$, we randomly generate a rotation matrix $\Rtrue\in \SO3$ and compute its rotated version $V_2\in\mathbb{R}^{L\times L\times L}$ using the Fourier transform-based approach mentioned above.  
The goal is then to recover the rotation $\Rtrue$ given only $V_1$ and $V_2$. In this subsection we focus on pure rotation recovery and incorporation of translation will be demonstrated in Section \ref{sec:numerics-compare}. The test volumes shown in Figure \ref{figure:volumes} will be used throughout the numerical experiments.

\subsubsection{Algorithm \ref{algo:bo align} without Refinement}
As mentioned, we shall first downsample the given volumes to $V_i^{\text{\tiny DS}}\in\mathbb{R}^{L_0\times L_0 \times L_0}$ and then align the $V_i^{\text{\tiny DS}}$'s with Algorithm \ref{algo:bo align}. 
The downsampling level $L_0$ and the total number of iterations $T$ in Algorithm \ref{algo:bo align} are treated as user-chosen parameters which could vary depending on the molecules at hand. Below we shall investigate their effects on the performance of Algorithm \ref{algo:bo align}, first without the refinement step to focus on the Bayesian optimization performance. 

\begin{figure}[!htb]
\minipage{0.25\textwidth}
\includegraphics[width=\textwidth]{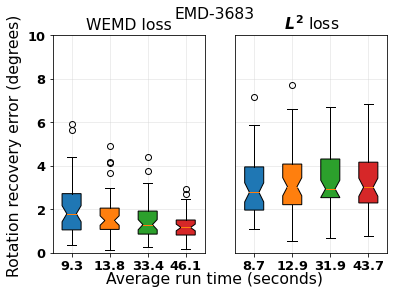}
\endminipage
\minipage{0.25\textwidth}
\includegraphics[width=\textwidth]{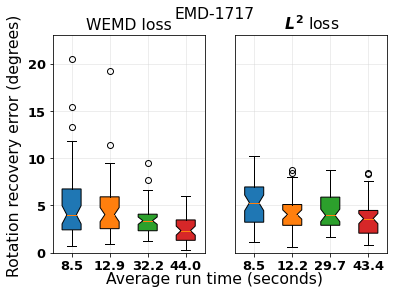}
\endminipage
\minipage{0.25\textwidth}
\includegraphics[width=\textwidth]{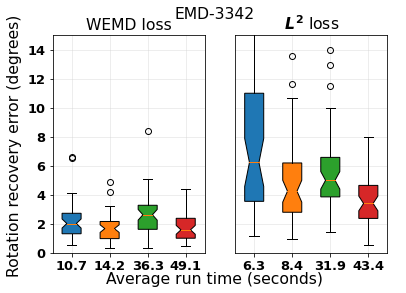}
\endminipage
\minipage{0.25\textwidth}
\includegraphics[width=\textwidth]{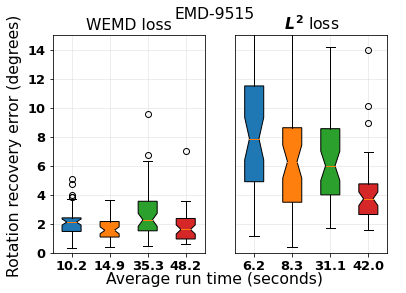}
\endminipage

\minipage{0.25\textwidth}
\includegraphics[width=\textwidth]{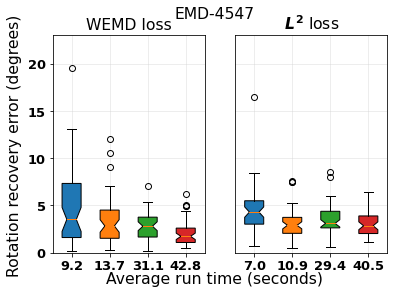}
\endminipage
\minipage{0.25\textwidth}
\includegraphics[width=\textwidth]{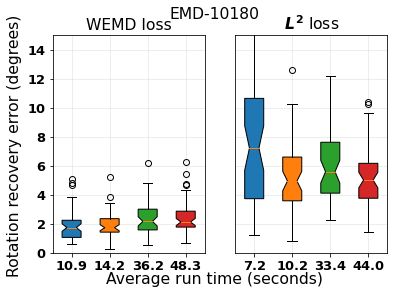}
\endminipage
\minipage{0.25\textwidth}
\includegraphics[width=\textwidth]{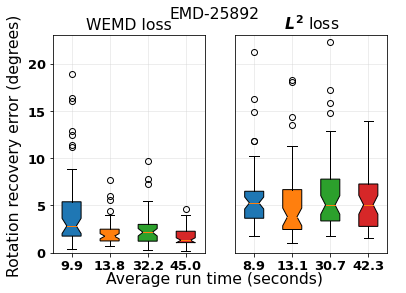}
\endminipage
\minipage{0.25\textwidth}
\includegraphics[width=\textwidth]{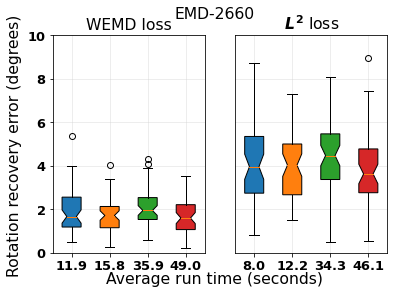}
\endminipage
\caption{Performance comparison between Algorithm \ref{algo:bo align} and its $L^2$ loss version without refinement. The four boxplots in each subfigure correspond to (from left to right) $(L_0,T)=(32,150)$, $(32,200)$, $(64,150)$, $(64,200)$. The vertical axis represents rotation recovery error $|\Theta(\Rtrue,\RBO)|$ in degrees. The tick labels record the average run time in seconds}\label{figure:BO}
\end{figure}

Denoting the estimated rotation by $\RBO$, we quantify the performance by the relative angle 
$|\Theta(\Rtrue,\RBO)|$ between $\Rtrue$ and $\RBO$ defined in \eqref{eq:cf geo}. Figure \ref{figure:BO} shows the results for four combinations of $L_0$ and $T$ for the molecules shown in Figure \ref{figure:volumes}. 
To illustrate the benefits of alignment in Wasserstein distances, also shown in Figure \ref{figure:BO} are results for the parallel experiments with WEMD loss in Algorithm \ref{algo:bo align} replaced by the $L^2$ loss. 
The experiments are repeated 50 times with $\Rtrue$ regenerated in each. The run time is recorded on a laptop with Intel(R) Core(TM) i7-7500 CPU@ 2.70GHz. We see that with high probability, the WEMD version of Algorithm \ref{algo:bo align} is able to recover the relative rotation up to a 5-degree error with only 200 evaluations of the loss function and in most cases outperforms its $L^2$ counterpart with comparable computing time.

\subsubsection{Algorithm \ref{algo:bo align} with Refinement}
With the results shown in Figure \ref{figure:BO}, we shall continue to demonstrate the performance of Algorithm \ref{algo:bo align} when refinement is included. We recall that the refinement step is a local search \eqref{eq:refine sol} solved with the Nelder-Mead algorithm initialized at the estimate $\RBO$ returned by the first part of Algorithm \ref{algo:bo align}. 
We note that here we have the freedom to choose the combination of $L_0$ and $T$ for obtaining $\RBO$ with optimal efficiency. 
In particular, setting $L_0=64$ and $T=200$ as shown in Figure \ref{figure:BO} leads to the best overall performance, but the other choices also give reasonably accurate recovery while requiring much less run time. 
Meanwhile, it would not be too surprising that the $\RBO$'s returned by these other choices could lie in the basin of attraction around $\Rtrue$ so that local convergence is still retained after the refinement.

\begin{figure}[!htb]
\minipage{0.25\textwidth}
\includegraphics[width=\textwidth]{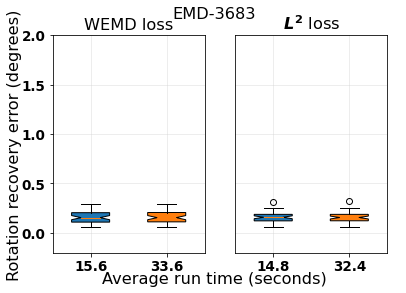}
\endminipage
\minipage{0.25\textwidth}
\includegraphics[width=\textwidth]{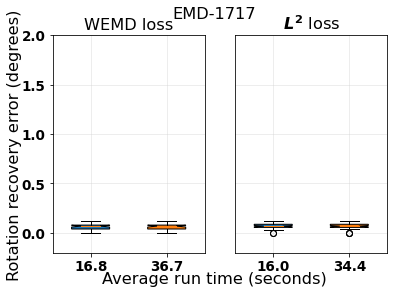}
\endminipage
\minipage{0.25\textwidth}
\includegraphics[width=\textwidth]{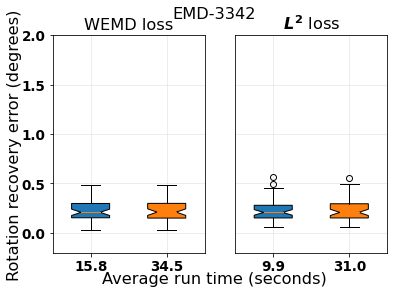}
\endminipage
\minipage{0.25\textwidth}
\includegraphics[width=\textwidth]{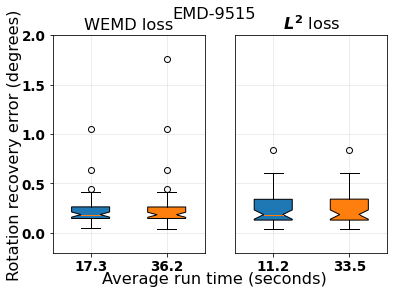}
\endminipage

\minipage{0.25\textwidth}
\includegraphics[width=\textwidth]{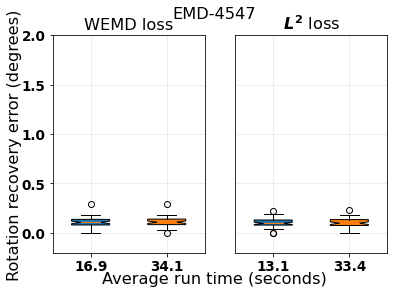}
\endminipage
\minipage{0.25\textwidth}
\includegraphics[width=\textwidth]{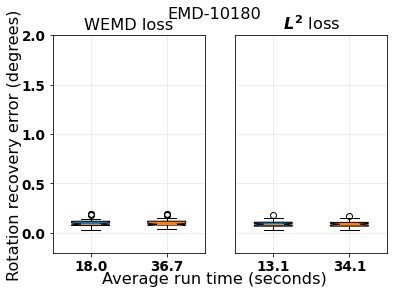}
\endminipage
\minipage{0.25\textwidth}
\includegraphics[width=\textwidth]{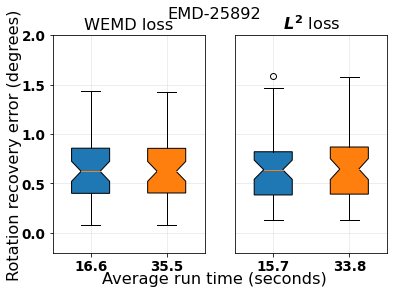}
\endminipage
\minipage{0.25\textwidth}
\includegraphics[width=\textwidth]{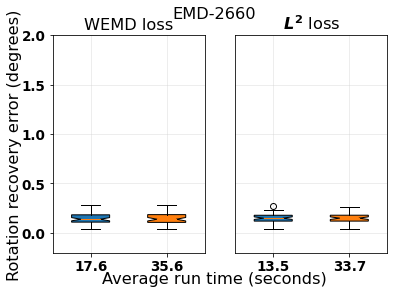}
\endminipage
\caption{Performance comparison between Algorithm \ref{algo:bo align} and its $L^2$ loss version with refinement. The two boxplots in each subfigure correspond to (from left to right) $(L_0,T)=(32,200)$ and $(64,150)$. The vertical axis represents rotation recovery error $|\Theta(\Rtrue,\RBO)|$ in degrees. The tick labels record the average run time in seconds}\label{figure:BO-R}
\end{figure}

In Figure \ref{figure:BO-R}, we show that this is indeed the case for the combinations $(L_0,T)=(32,200)$ and $(64,150)$ for both the WEMD and $L^2$ versions of Algorithm \ref{algo:bo align}, which give after refinement more accurate recovery but with less run time than using the combination $(64,200)$ without refinement. 
Here in the refinement step, we are fixing the downsampling level to be 32.   
In particular, we see that the local search step by Nelder-Mead appears to be not very stringent on its initializations so that even those returned by the $L^2$ version of Algorithm \ref{algo:bo align} would suffice for good final accuracy. However, this could be a coincidence due to the benign structures of the test volumes. The better initial estimates returned by the WEMD version as shown in Figure \ref{figure:BO} could already be 
returned as a solution and at the same time are more reassuring as initializations for the local refinement. For this reason, the WEMD version  serves as our main algorithm.

\subsubsection{Robustness to Noise}

\begin{figure}[!htb]
\minipage{0.125\textwidth}
\subcaption*{\scriptsize \,\, SNR=1/4}
\includegraphics[width=\textwidth]{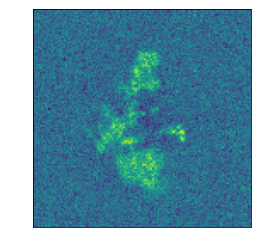}
\endminipage
\minipage{0.125\textwidth}
\subcaption*{\scriptsize \,\, SNR=1/8}
\includegraphics[width=\textwidth]{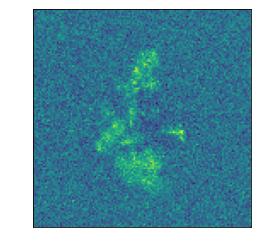}
\endminipage
\minipage{0.125\textwidth}
\subcaption*{\scriptsize \,\, SNR=1/16}
\includegraphics[width=\textwidth]{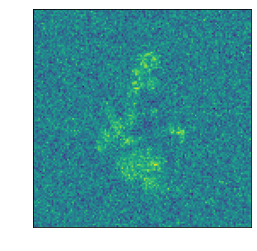}
\endminipage
\minipage{0.125\textwidth}
\subcaption*{\scriptsize \,\, SNR=1/32}
\includegraphics[width=\textwidth]{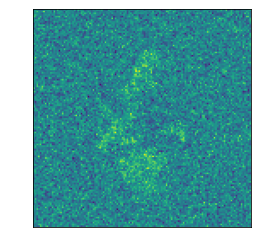}
\endminipage
\minipage{0.125\textwidth}
\subcaption*{\scriptsize \,\, SNR=1/64}
\includegraphics[width=\textwidth]{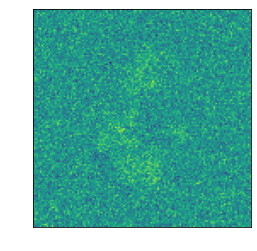}
\endminipage
\minipage{0.125\textwidth}
\subcaption*{\scriptsize \,\, SNR=1/128}
\includegraphics[width=\textwidth]{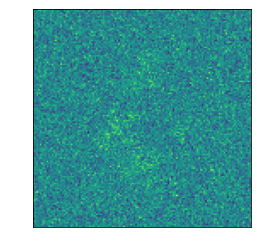}
\endminipage
\minipage{0.125\textwidth}
\subcaption*{\scriptsize \,\, SNR=1/256}
\includegraphics[width=\textwidth]{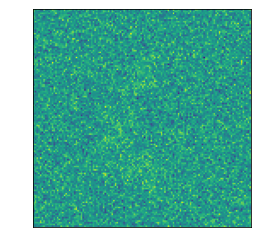}
\endminipage
\minipage{0.125\textwidth}
\centering
\subcaption*{\scriptsize \,\, SNR=1/512}
\includegraphics[width=\textwidth]{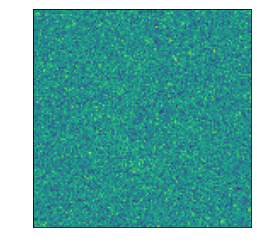}
\endminipage
\caption{Visualization of a central slice of EMD-3683 under different signal-to-noise ratios}\label{figure:noise}
\end{figure}

\begin{figure}[!htb]
\minipage{0.25\textwidth}
\includegraphics[width=\textwidth]{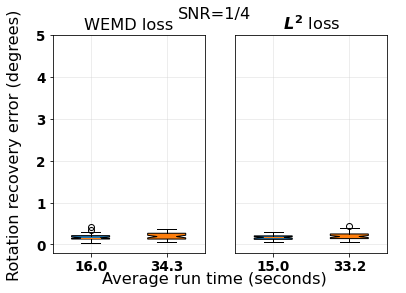}
\endminipage
\minipage{0.25\textwidth}
\includegraphics[width=\textwidth]{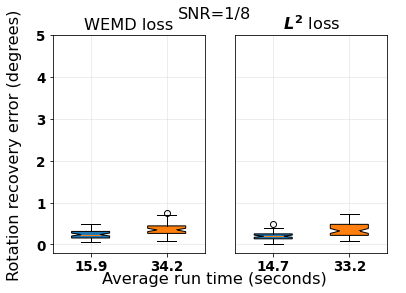}
\endminipage
\minipage{0.25\textwidth}
\includegraphics[width=\textwidth]{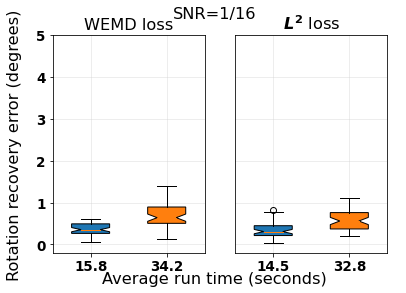}
\endminipage
\minipage{0.25\textwidth}
\includegraphics[width=\textwidth]{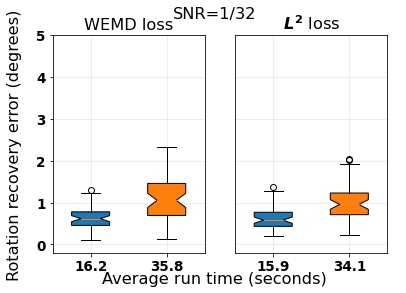}
\endminipage

\minipage{0.25\textwidth}
\includegraphics[width=\textwidth]{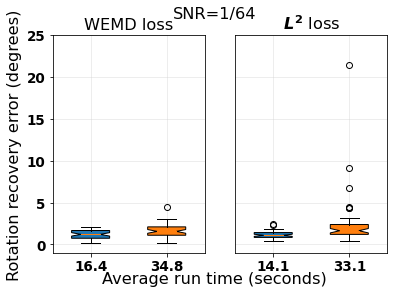}
\endminipage
\minipage{0.25\textwidth}
\includegraphics[width=\textwidth]{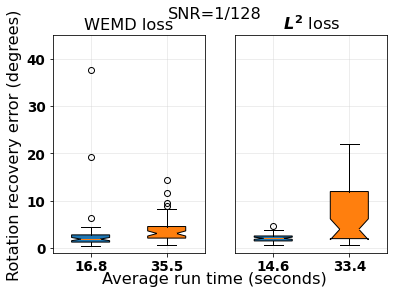}
\endminipage
\minipage{0.25\textwidth}
\includegraphics[width=\textwidth]{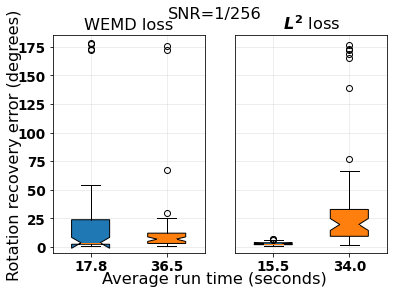}
\endminipage
\minipage{0.25\textwidth}
\includegraphics[width=\textwidth]{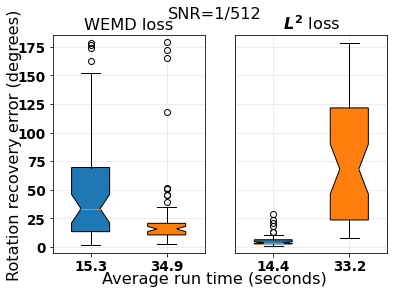}
\endminipage
\caption{Performance comparison between Algorithm \ref{algo:bo align} and its $L^2$ loss version with refinement under noise corruption. The two boxplots in each subfigure correspond to (from left to right) $(L_0,T)=(32,200)$ and $(64,150)$. The vertical axis represents rotation recovery error $|\Theta(\Rtrue,\RBO)|$ in degrees. The tick labels record the average run time in seconds}\label{figure:BO-noise}
\end{figure}

We further test the performance of Algorithm \ref{algo:bo align} in the presence of noise, where we fix the test volume to be EMD-3683 and add to each entry of $V_1,V_2$  an independent Gaussian noise of variance $\sigma^2$ across a range of signal-to-noise ratios, defined as SNR=$\|V_1\|_2^2/(L^3\sigma^2)$. 
Figure \ref{figure:noise} visualizes a central slice of the noise corrupted volumes. 
The performance of Algorithm \ref{algo:bo align} is shown in Figure \ref{figure:BO-noise}, which shows a decent level of robustness.
An interesting observation is that the $L^2$ version with downsampling level $L_0=32$ appears to be more robust than the WEMD version for very high noise levels. This is related to our discussion in Section \ref{sec:heter} on the alignment of heterogeneous pairs since the noise corrupted volumes can be treated as different conformations of the clean one. We will show by an example that the Wasserstein-based distances could be more susceptible to heterogeneity.

\subsection{Comparison with Existing Algorithms} \label{sec:numerics-compare}        
Finally, we shall compare Algorithm \ref{algo:bo align} with two recent alignment algorithms proposed by \cite{harpaz2023three}, which exploits common line-based methods for fast projection matching, and \cite{riahi2023alignot}, which considers an entropic regularization of 2-Wasserstein loss in \eqref{eq:alignment problem rot W} and employs stochastic gradient descent for optimization. In the following comparison, we shall also consider translation recovery. More precisely, given a volume $V_1\in\mathbb{R}^{L\times L\times L}$, we randomly generate a rotation matrix $\Rtrue$ and first compute its rotated version $V_2^{\text{\tiny rot}}$, whose shifted version $V_2$ is then treated as the given volume. Here the shift vector is uniformly randomly generated over the cube $[-S,S]^3$ with $S=0.05L$. This corresponds to a typical situation in cryo-EM applications where the given volumes are already preprocessed and approximately centered. 

As mentioned in the introduction, we shall recover the shift by centering the volumes. We point out that the volumes given in the database \cite{lawson2016emdatabank} contain negative values so a thresholding is applied first before computing the center of mass, where the threshold is chosen based on the recommended contour level for each molecule in \cite{lawson2016emdatabank} or could be tuned empirically. The code provided by \cite{riahi2023alignot} also focuses only on rotation recovery so we apply the same centering step for their algorithm.

\begin{figure}[!htb]
\minipage{0.25\textwidth}
\includegraphics[width=\textwidth]{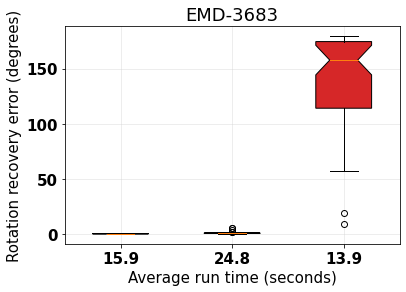}
\endminipage
\minipage{0.25\textwidth}
\includegraphics[width=\textwidth]{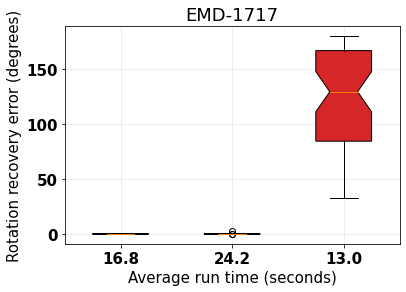}
\endminipage
\minipage{0.25\textwidth}
\includegraphics[width=\textwidth]{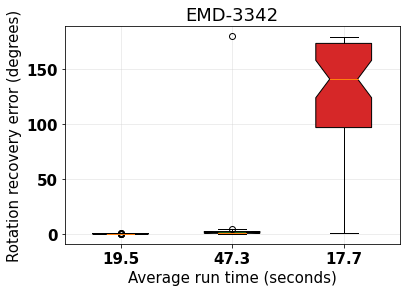}
\endminipage
\minipage{0.25\textwidth}
\includegraphics[width=\textwidth]{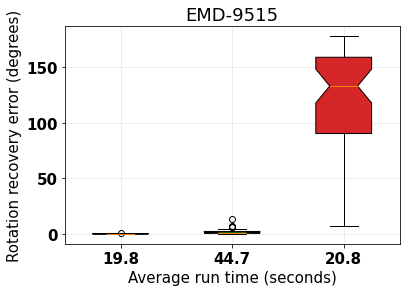}
\endminipage

\minipage{0.25\textwidth}
\includegraphics[width=\textwidth]{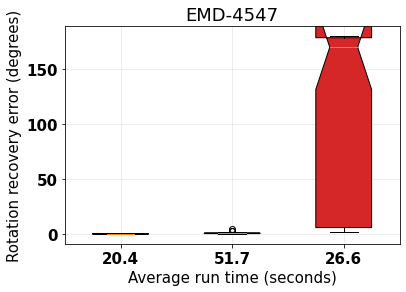}
\endminipage
\minipage{0.25\textwidth}
\includegraphics[width=\textwidth]{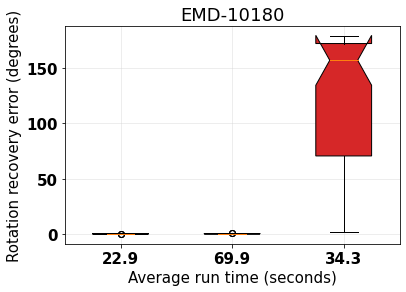}
\endminipage
\minipage{0.25\textwidth}
\includegraphics[width=\textwidth]{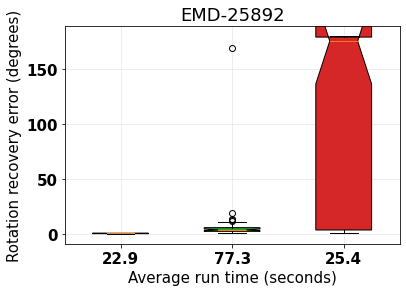}
\endminipage
\minipage{0.25\textwidth}
\includegraphics[width=\textwidth]{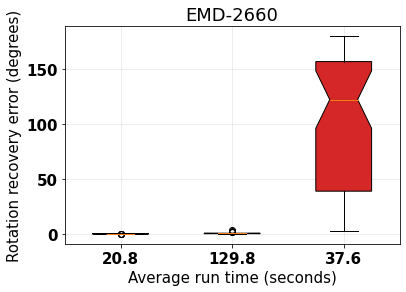}
\endminipage
\caption{Comparison with existing methods. The three boxplots in each subfigure correspond to (from left to right) BOTalign (our method), EMalign, and AlignOT. The vertical axis represents rotation recovery error $|\Theta(\Rtrue,\RBO)|$ in degrees. The tick labels record the average run time in seconds}\label{figure:compare}
\end{figure}

For the comparisons below, we shall use $(L_0,T)=(32,200)$ with refinement in our Algorithm \ref{algo:bo align}, which will be denoted as Bayesian Optimal Transport Align (BOTalign). The algorithm in \cite{harpaz2023three} will be denoted as EMalign following the authors, and is applied with downsampling 32 and their recommended number of reference projections 30, also implemented with their local refinement. Lastly, AlignOT stands for the algorithm in \cite{riahi2023alignot} with $n=500$ in their topology representing network step and maximum number of iteration $N=500$ in their stochastic gradient descent. Again, we repeat the experiments 50 times for each molecule and the results are shown in Figure \ref{figure:compare}   and Table \ref{tab:compare}.

\begin{table}[!htb]
    \centering
    \begin{tabular}{|c|c|c|c|c|c|c|c|c|}
    \hline
        EMD ID
         & 3683 & 1717 & 3342 & 9515 & 4547 & 10180 & 25892 & 2660\\ 
         \hline  
         \multirow{2}{4em}{BOTalign}
         & 0.26 & 0.11 & 0.23 & 0.25 & 0.11 & 0.10 & 0.58 & 0.18\\ 
         & (0.13) & (0.05) & (0.11) & (0.14) & (0.05) & (0.04) & (0.29)  & (0.08)\\ 
         \hline  
         \multirow{2}{4em}{EMalign}
         & 0.95 & 0.15 & 5.18 & 2.04 & 0.74 & 0.24 & 8.09 & 0.60\\ 
         & (1.09) & (0.31) & (24.98) & (2.23) & (0.77) & (0.18) & (23.34) & (0.60)\\
         \hline  
         \multirow{2}{4em}{AlignOT}
         & 140.73 & 122.12 & 123.53 & 119.26 & 97.92 & 118.00 & 126.69 & 103.96\\
         & (43.33) & (45.87)  & (56.28) & (49.13) & (85.39) & (69.14) & (77.90) & (61.38) \\
         \hline 
    \end{tabular}
    \caption{Summary statistics of the recovery errors in Figure \ref{figure:compare}. In each block are the mean and standard deviation (in parentheses). }
    \label{tab:compare}
\end{table}

We see that our algorithm achieves the best accuracy with minimal run time. We remark that AlignOT is a local search algorithm where the authors report good recovery if the angle between $\Rtrue$ and the identity is within 75 degrees. This is not contradictory with the results shown in Figure \ref{figure:compare} as here $\Rtrue$ is randomly generated, which more than half of the time is 75 degrees away from the identity.  
EMalign is a global search algorithm that improves over earlier alignment algorithms especially in terms of running time (see their Tables 2 and 3). Our algorithm achieves further improvements when aligning clean molecules. We mention that EMalign performs exhaustive search over shifts as well as rotations and could have an advantage when the noise level is high as the centering step would be less accurate. 
In this case, more sophisticated center of mass estimation method such as \cite{heimowitz2021centering} needs to be employed instead in our approach.

\subsection{Algorithmic Complexity}\label{sec:complexity}
Here we briefly discuss the complexity of our Algorithm \ref{algo:bo align} with respect to the size $L$ of the volumes. The majority of the computational cost of our algorithm takes place in (i) solving the surrogate problem \eqref{eq:pick candiate} and (ii) evaluating the loss $\Fwemd$. The former step is always a three-dimensional optimization problem and does not depend explicitly on $L$. Our empirical experience suggests that with the early stopping that we have adopted its cost is relatively small compared to evaluating $\Fwemd$. The latter would involve two steps, where one first rotates one of the volumes using nonuniform fast Fourier transform that costs $O(L^3\log L)$, and then computes the WEMD with a cost of $O(L^3)$.
  One may need to use a larger maximum scale level $s$ for large $L$'s but the dependence is in general only logarithmic.   
Therefore the total cost of Algorithm \ref{algo:bo align} is on the order of $O(TL^3 \log L)$ where $T$ is the total number of iterations which can be fixed for instance as 200. Therefore the dependence on $T$ improves over naive exhaustive search methods which cost $O(|S|\times L^3)$ where $|S|=O(10^5)\sim O(10^6)$ is the size of the rotation grid to search over, and the dependence on $L$ improves over the convolution based methods such as \cite{kyatkin2000algorithms,chen2013fast} which would take $O(L^4)$.

\section{Alignment of Heterogeneous Pairs} \label{sec:heter}

In this section, we shall discuss the problem of aligning a pair of similar but non-identical volumes, which we denote as a heterogeneous pair. Such problem arises naturally in cryo-EM, where the same protein molecule can exhibit different conformations. For example, a molecule can consist of two moving parts that are rotating with respect to each other, or the molecule can be more elongated in certain states than others. We shall show by an example that alignment in $W_1$ distance as we have proposed could be problematic in the presence of heterogeneity. The same issue is present for the $L^2$ distance although milder. This motivates a need of new loss functions for aligning a heterogeneous pair, which ideally extracts and compares the common part of volumes. Such sophisticated loss functions are likely to render gradient-based or exhaustive search-based optimization ineffective, whereas our algorithmic framework in Algorithm \ref{algo:bo align} could be seamlessly incorporated.

\begin{figure}[!htb]
    \centering  
    \vspace{-0.4cm}
    \includegraphics[width=0.75\linewidth]{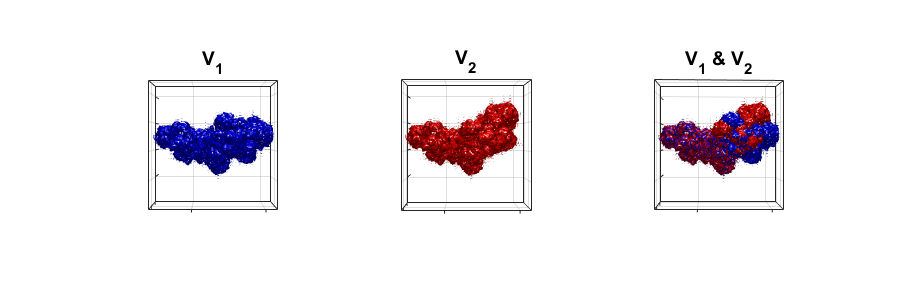}
    \vspace{-1cm}
    \caption{Side views of a heterogeneous pair of volumes. Notice that $V_1$ and $V_2$ differ mostly in the upper right portion. } \label{figure:heter-merge}
\end{figure}

\begin{figure}[!htb]
    \centering  
    \vspace{-0.4cm}
    \includegraphics[width=0.75\linewidth]{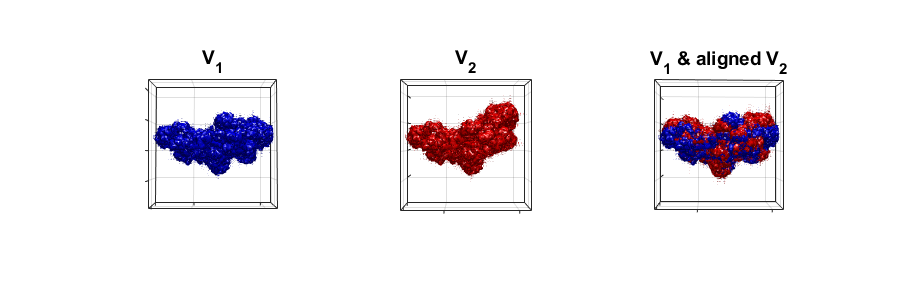}
    \vspace{-1cm}
    \caption{Volume alignment with heterogeneity in Wasserstein distance} \label{figure:exp-heter}
\end{figure}

  To start with, let's consider a pair of simulated volumes $V_1$ and $V_2$ in Figure \ref{figure:heter-merge}, which have an overall similar shape but certain differences in their upper right parts, representing different conformations of the same molecule. The two volumes are considered as already aligned as the ``base'' portions of the volumes are matched.    However, if we apply Algorithm \ref{algo:bo align} (without refinement) to this pair, the resulting aligned volume is shown as the rightmost subplot in Figure \ref{figure:exp-heter}, which turns out to be misaligned,   in that the left portion of the volumes are now mismatched.   

\begin{figure}[!htb]
\centering   
\vspace{0.2cm}
    \includegraphics[width=0.5\linewidth]{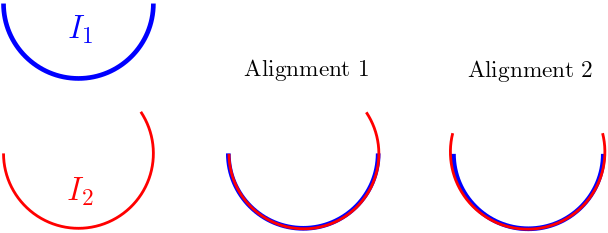}
    \vspace{0.2cm}
    \caption{Synthetic hetergeneous pair} \label{figure:example-heter}
\end{figure}

To see what is happening in this case, let's give a heuristic explanation by considering the following two-dimensional abstraction of the volumes in Figure \ref{figure:example-heter}. Here $I_1$ is a semi-circle and $I_2$ is an extended (but slightly thinner) semi-circular arc with the same radius as $I_1$, where the extended portion of $I_2$ is considered as the heterogeneous part, in analogy with $V_1$ and $V_2$ in Figure \ref{figure:exp-heter}. Now we consider two possible alignment of these two arcs as in Figure \ref{figure:example-heter} and give a rough calculation of the corresponding $W_1$ losses in both cases. 

Suppose the extended portion in $I_2$ has length $\epsilon$. Recall that the $W_1$ distance has the interpretation of the amount of mass transportation needed from arc $I_1$ to the arc $I_2$.
For Alignment 1, the optimal transport plan would be to map both the endpoints of $I_1$ to the endpoints of $I_2$, and to map every other point in the middle in a proportional way. This would incur a $W_1$ loss approximately equal to 
\begin{align}
    \tag{Alignment 1}
    W_1(I_1,I_2)\approx \int_0^1\epsilon x \,dx=\frac{\epsilon}{2}.
\end{align}
For Alignment 2, the optimal transport plan would be to map each half of $I_1$ proportionally to the corresponding halves of $I_2$, which would incur a $W_1$ loss roughly 
\begin{align}
    \tag{Alignment 2}
    W_1(I_1,I_2)\approx 2\cdot \int_{0}^{1/2}\epsilon x\,dx=\frac{\epsilon}{4}.
\end{align}
Therefore Alignment 2 leads to a smaller $W_1$ loss and is preferred as we have observed in Figure \ref{figure:exp-heter}. 

\begin{figure}[!htb]
\centering   
\vspace{-0.4cm}
    \includegraphics[width=0.75\linewidth]{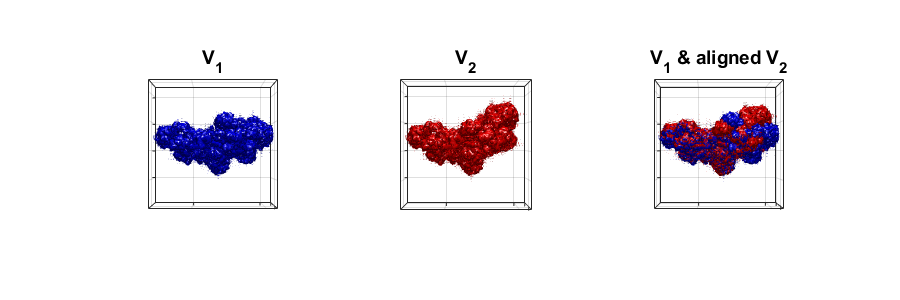}
     \vspace{-1cm}
    \caption{Volume alignment with heterogeneity in Euclidean distance} \label{figure:example-heter-eu}
\end{figure}

On the other hand, the $L^2$ losses between $I_1$ and $I_2$ remain the same regardless of whether Alignment 1 or 2 is applied.  An implementation of Algorithm \ref{algo:bo align} with WEMD replaced by the $L^2$ loss actually gives close-to-correct alignment as shown in Figure \ref{figure:example-heter-eu}, with an error about 3 degrees.   This suggests that the $L^2$ loss might be a better choice in the presence of heterogeneity   and explains its choice in our refinement step to avoid the effect of perturbations of the given volumes. 

However, the empirical success of $L^2$ alignment in Figure \ref{figure:example-heter-eu} may be a coincidence thanks to other finer structures of the molecules. 
Our synthetic example in Figure \ref{figure:example-heter} does raise an interesting question on what loss function to be used when aligning heterogeneous pairs. 
In particular, there is a potential need of a more sophisticated distance function $d_{\text{heter}}$ that for instance only compares the shared components in the volumes by extracting the common features. 
  In the context of compositional heterogeneity where the two volumes have different total masses as a result of missing subunits, \cite{riahi2023empot} proposes a partial alignment procedure based on Gromov-Wasserstein divergence. 
We shall leave such investigation of more advanced distance functions for future work.  
Meanwhile, we remark that in this case, exhaustive search-based methods that rely on efficient representations of the loss function over $\SO3$ such as \cite{chen2013fast} would be challenging unless certain special structures of $d_{\text{heter}}$ can be exploited. 
The same is likely to be true for gradient-based methods where one would need to rely on numerical differentiation which could be less accurate and inefficient. 
Nevertheless, our Algorithm \ref{algo:bo align} provides a ready-to-use recipe as long as we can evaluate $d_{\text{heter}}$.

\section{Discussion}
In this paper we proposed an alignment algorithm using a Wasserstein-based distance as the loss function, which is optimized with tools from Bayesian optimization followed by a local refinement procedure. Numerical experiments show improved performance of our algorithm over existing methods on alignment of real protein molecules from cryo-EM. The proposed algorithm can be extended to arbitrary loss functions, which could be a feasible solution in the presence of heterogeneity where we have illustrated a potential need of novel distance functions. We have presented the algorithmic framework for volumes represented as density maps, but it can be easily extended to other volume representations as long as one can define a suitable loss function as in \eqref{eq:alignment problem rot}.

Our algorithm focuses mainly on the clean volume case, where the problem reduces to rotation estimation since the relative translation can be recovered by the centering step. 
We remark that this is not a simplifying assumption in the context of cryo-EM since 3D alignment is usually carried out on reconstructed molecules, which are much cleaner than their projection images to start with. 
However, this could be a limitation of our approach, where in the presence of noise more sophisticated center of mass estimation methods such as \cite{heimowitz2021centering} would be necessary.  

On the other hand, we remark that it is possible to incorporate translation estimation in the Bayesian optimization framework that we have adopted. In particular, the alignment objective in \eqref{eq:alignment problem rot wemd} and its surrogate problem \eqref{eq:pick candiate} can be both extended to the product space $\mathbb{B}\times \SO3$. The algorithmic framework proceeds as before except the covariance function defining the GP would also need to be extended to  $\mathbb{B}\times \SO3$. This for instance can be achieved simply with the product covariance
\begin{align*}
    c\big((v,R),(w,S)\big)= \sigma^2 \exp\left(-\frac{\|R-S\|_F^2}{2\ell_r^2}\right) \exp\left(-\frac{\|v-w\|_2^2}{2\ell_s^2}\right). 
\end{align*}
However, our simulations suggest slow exploration of the search space and poor recovery in comparable amount of time as existing methods. This may be due to the fact that Bayesian optimization is known to work better in lower dimensions. The additional set $\mathbb{B}$ doubles the search space dimension and necessitates many more observations of the loss function to decode its landscape.

  In the context of aligning heterogeneous pairs of volumes, there are still many improvements of the current work that are worth exploring. As a referee pointed out, the WEMD framework might be able to inform us about which parts of the volumes need further transport after obtaining an optimal alignment. Here we outline the idea without going into the mathematical details. 
As noted in \cite{shirdhonkar2008approximate}, computing the $W_1$ distance between two density maps $\phi_1$ and $\phi_2$ is equivalent to the following optimization problem 
\begin{align*}
    \underset{f \text{ 1-Lipschitz}}{\operatorname{max}}\,\,  \int f(x) [\phi_1(x)-\phi_2(x)] dx,
\end{align*}
whose solution $f$ can be approximated by a wavelet expansion with coefficients $f_\lambda= \text{sign}(p_\lambda)\cdot 2^{-j(1+n/2)}$, where $p_\lambda$'s are the wavelet coefficients of the difference density map $p=\phi_1-\phi_2$. 
Therefore $f$ is computable and according to the Kantorovich-Rubenstein duality theorem, it corresponds to the Lagrange multiplier of the marginal constraints involving $\phi_1$ and $\phi_2$. 
It can then be used to locate regions where the marginal constraints could be relaxed and hence regions where further transport is needed. 
This would have important applications in heterogeneous alignment and we wish to explore such ideas in the future.

\section*{Acknowledgements}
The authors are supported in part by AFOSR under Grant FA9550-20-1-0266, in part by Simons Foundation Math+X Investigator Award, in part by NSF under Grant DMS-2009753, and in part by NIH/NIGMS under Grant R01GM136780-01.
R.Y. would like to thank 
Marc Aurèle Gilles and Oscar Mickelin for sharing protein data and useful codes.

\bibliographystyle{abbrvnat} 
\bibliography{bib}   

\end{document}